\documentclass[aps,pre,amsmath,amssymb,twocolumn,showpacs,superscriptaddress]{revtex4}
\usepackage[usenames,dvipsnames]{color}
\usepackage{graphicx}

\newcommand{\avg}[1]{\langle #1 \rangle}
\newcommand{\ahum}[1]{``#1''}
\newcommand{\eq}[1]{Eq.~(\ref{#1})}
\newcommand{\fig}[1]{Fig.~\ref{#1}}

\newcommand{\tab}[1]{Table~\ref{#1}}
\newcommand{\olcite}[1]{Ref.~\citenum{#1}}

\newcommand{\rhoc}{\rho_{\rm cr}}
\newcommand{\rhom}{\rho_M}
\newcommand{\zcr}{z_{\rm cr}}

\begin{document}

\title{Fluids in porous media: The case of neutral walls}

\author{Giuseppe Pellicane}
\affiliation{School of Chemistry and Physics, University of Kwazulu-Natal, Private Bag X01, Scottsville 3209,
Pietermaritzburg, South Africa 
}
\affiliation{National Institute for Theoretical Physics (NITheP), KZN node, Pietermaritzburg, South Africa}

\author{Richard L. C. Vink} 
\affiliation{Institute of Theoretical Physics, Georg-August-Universit\"at, 
Friedrich-Hund-Platz 1 D-37077 G\"ottingen, Germany}

\author{Bruno Russo}
\affiliation{Techimp Corporate Headquarter, Via Toscana, 11/c, 40069 Zola Predosa (BO), Italy}

\author{Paolo V. Giaquinta}
\affiliation{Dipartimento di Fisica e di Scienze della Terra, Universit\`a degli Studi di Messina, 
Viale F. Stagno d'Alcontres 31, 98166 Messina, Italy}

\date{\today}

\begin{abstract} The bulk phase behavior of a fluid is typically altered when 
the fluid is brought into confinement by the walls of a random porous medium. 
Inside the porous medium, phase transition points are shifted, or may disappear 
altogether. A crucial determinant is how the walls interact with the fluid 
particles. In this work, we consider the situation whereby the walls are neutral 
with respect to the liquid and vapor phase. In order to realize the condition of 
strict neutrality, we use a symmetric binary mixture inside a porous medium that 
interacts identically with both of the mixture species. Monte Carlo simulations 
are then used to obtain the phase behavior. Our main finding is that, in the 
presence of the porous medium, a liquid-vapor type transition still occurs, but 
with critical exponents that deviate from bulk Ising values. In addition, we 
observe clear violations of self-averaging. These findings provide further 
evidence that random confinement by neutral walls induces critical behavior of 
the random Ising model (i.e.~Ising models with dilution type disorder, where the 
disorder couples to the energy). \end{abstract}



\maketitle

\section{Introduction}

The confinement of a fluid to the voids of a porous material generally 
influences the critical behavior of the fluid. For example, lutidine-water 
mixtures in Vycor~\cite{wiltzius:1989}, or $^{4}$He~\cite{wong.chan:1990}, 
nitrogen~\cite{wong.kim.ea:1993} and carbon dioxide~\cite{melnichencko:2004} in 
silica aerogel, yield critical exponents of their associated liquid-vapor 
transitions that differ profoundly from bulk values (the bulk exponents 
typically being those of the three-dimensional Ising model). One line of thought 
is that the random pore structure induces quenched spatial fluctuations in the 
chemical potential~\cite{maher:1984}. This conjecture, originally put forward by 
de~Gennes~\cite{gennes:1984}, implies that the critical behavior of the fluid 
inside the pores should be that of the random-field Ising model 
(RFIM)~\cite{nattermann:1998, citeulike:10004339}. Recent simulations of fluids 
inside porous media have indeed uncovered critical behavior characteristic of 
the RFIM~\cite{vink.binder.ea:2006, vink.binder.ea:2008, citeulike:9736167, 
citeulike:8864903}. In order for RFIM universality to arise, it is crucial that 
the pore walls feature a preferred attraction to one of the fluid phases. This 
condition is typically fulfilled in experiments, as one of the phases, i.e.~the 
liquid or the vapor, is frequently seen to wet the pore 
walls~\cite{wiltzius:1989, wong.kim.ea:1993, melnichencko:2004}.

Nevertheless, for our fundamental understanding of fluid phase behavior, the 
situation of \ahum{neutral} pore walls which do not preferentially attract, is 
of interest also. A different universality class is then expected to come into 
play~\cite{kierlik.pitard.ea:1998, lucentini.pellicane:2008}, namely the one of 
the random Ising model (RIM). The defining feature of the RIM is that the 
quenched randomness of the porous medium couples to the energy (as opposed to 
the order parameter in the RFIM). Typical lattice models that belong to the 
universality class of the RIM are the site-diluted Ising model and the 
random-bond Ising model~\cite{hasenbusch.toldin:2007}. In $d=3$ dimensions, the 
Harris criterion~\cite{harris:1974} implies that the RIM should still feature a 
liquid-vapor critical point, but with critical exponents different from those of 
the bulk Ising model, since the latter has a positive specific heat critical 
exponent (by bulk we mean in the absence of the porous medium). However, the 
difference in the critical exponents between bulk Ising and RIM universality is 
very small, and challenging to detect numerically~\cite{hasenbusch.toldin:2007}. 
In contrast, the difference between bulk Ising and the RFIM is much more 
pronounced, since hyperscaling is violated in the latter. In $d=3$ dimensions, 
this yields a very pronounced numerical signature which one can easily detect in 
simulations~\cite{vink.binder.ea:2006, vink.binder.ea:2008, citeulike:9736167, 
citeulike:10004339}.

Regarding the case of a fluid confined to a neutral porous medium, the question 
of whether this system exhibits RIM universality was recently addressed in 
simulations~\cite{lucentini.pellicane:2008}. As expected for the RIM, these 
simulations revealed a critical point, located at an increased density compared 
to the bulk. By carefully measuring the critical amplitude ratio of the 
susceptibility, these simulations also uncovered deviations from bulk Ising 
behavior, and toward that of the RIM. The aim of this work is to corroborate 
these findings, using a more sophisticated (grand-canonical) simulation scheme, 
larger system sizes, as well as additional finite-size scaling methods. In 
particular, we will address the question of self-averaging. The presented 
analysis provides further support of RIM universality in fluids confined to 
neutral pores.

The outline of this paper is as follows: In Section II, we introduce the model 
for the fluid mixture and for the porous medium with neutral walls, and we 
describe the simulation method. The results are presented in Section~III, and we 
end with a discussion in Section~IV.

\section{Model and Methods}

\subsection{Model: fluid inside neutral porous medium}
\label{sec:model}

We consider the same model as in \olcite{lucentini.pellicane:2008}, which is a 
fluid confined to a neutral porous medium in $d=3$ spatial dimensions. It 
belongs to the family of \ahum{quenched-annealed} 
mixtures~\cite{madden.glandt:1988, madden:1992}, which are routinely used to 
model fluids inside pores~\cite{gs, lomba.given:1993, page.monson:1996, 
kierlik.pitard.ea:1998, alvarez.levesque:1999, dps, pitard.rosinberg:1995, 
sarkisov.monson:2000, scholl-paschinger.levesque.ea:2001, pcwl, Sarkisov, 
vink.binder.ea:2006}. The fluid is a non-additive binary mixture of spheres, 
species $A$ and $B$, of equal diameter $\sigma$ (in what follows $\sigma$ is the 
unit of length). The particles interact via hard-sphere pair potentials
\begin{equation}
\label{eq:model}
\begin{split}
u_{AA}(r) = u_{BB}(r) = \begin{cases} 
  \infty & r<\sigma \\
  0      & \rm otherwise,
\end{cases} \\
u_{AB}(r) = \begin{cases}
  \infty & r< (1+\Delta) \sigma \\
  0      & \rm otherwise,
\end{cases} 
\end{split}
\end{equation}
with $r$ the center-to-center distance between a pair of particles, and $\Delta$ 
the non-additivity parameter. The porous medium is a fixed configuration of 
non-overlapping spheres, species $M$, also of diameter $\sigma$. These spheres 
are distributed randomly at the start of the simulation, at density $\rhom$, 
but remain immobile (quenched) thereafter. Only after the porous medium has been 
generated, are the (mobile) fluid particles inserted. Note that \eq{eq:model} is 
symmetric under the exchange of particle labels $A \leftrightarrow B$. In order 
to retain this symmetry, the medium particles $M$ interact symmetrically with 
the mobile fluid particles: $u_{AM}(r) = u_{BM}(r) \equiv u_{AA}(r)$. In this 
way, we ensure that the porous medium remains neutral, i.e.~does not 
preferentially attract one of the fluid species. As a consequence, we do {\it 
not} expect the critical behavior of the RFIM for this system.

For $\Delta>0$, the model of \eq{eq:model} exhibits a liquid-vapor type 
transition~\cite{gozdz:2003}. To analyze this transition later on, we introduce 
the overall fluid density $\rho=(N_A+N_B)/V$, and the composition (order 
parameter) 
\begin{equation}
\label{eq:op}
 m = (N_A - N_B) / V,
\end{equation}
where $N_\alpha$ is the number of particles of species $\alpha$, and $V$ the 
volume of the system. Provided $\rho>\rhoc$, two fluid phases are observed, I 
and II, characterized by a positive and negative composition, $m_I$ and 
$m_{II}$, respectively (due to symmetry $m_I=-m_{II}$). Precisely at 
$\rho=\rhoc$, the system becomes critical, where $m_I=m_{II}=0$. We emphasize 
that $\rhoc$ is not trivially known beforehand (its value depends on $\Delta$ 
and $\rhom$). For $\rho<\rhoc$, the system reveals only one phase. Of course, 
this behavior is analogous to that of the Ising model, if one identifies $m$ in 
\eq{eq:op} as the magnetization per spin~\cite{landau.binder:2000, 
wilding:2003}.

Our model is thus defined by the non-additivity parameter $\Delta$, and the 
density of the porous medium $\rhom$. In what follows, $\Delta=0.2$, while for 
the porous medium $\rhom=0.1$ and 0.2 will be considered, as well as the bulk 
situation $\rhom=0$. 

\subsection{Method: grand-canonical Monte Carlo}

Our simulations are performed in the grand-canonical (GC) ensemble, where the 
volume $V$ is constant, while the particle numbers $N_\alpha$ can fluctuate 
freely, as governed by the fugacity $z_\alpha$. Here, $\alpha \in \{A,B\}$ 
strictly refers to the mobile fluid, since the porous medium is quenched. Due to 
the symmetry of the model, it follows that $N_A=N_B$ at criticality, and so we 
set the particle fugacities equal: $z_A=z_B\equiv z$. The corresponding 
Boltzmann weight of a given particle configuration $w \propto z^{N_A+N_B} 
e^{-E/k_BT}$, with $E$ the potential energy given by \eq{eq:model}, $T$ the 
temperature, and $k_B$ the Boltzmann constant. Of course, for hard spheres, $T$ 
does not affect static equilibrium properties, and thus is irrelevant. The sole 
control parameter in our simulations is therefore the fugacity $z$. In this 
work, we use standard single particle Monte Carlo moves~\cite{frenkel.smit:2001} 
to generate particle configurations conform the weight~$w$. To enhance 
efficiency, histogram reweighting is used to extrapolate data obtained for one 
value of $z$ to different (nearby) values~\cite{ferrenberg.swendsen:1988}. The 
simulations are performed in a cubic box of edge $L$ with periodic boundaries. 

The principal output of the simulations is the (normalized) distribution
\begin{equation}
\label{eq:opd}
 P(m) \equiv P(m|z,L,\rhom) ,\quad
 \int_{-\infty}^{\infty} P(m) \, dm=1,
\end{equation}
defined as the probability to observe the system in a state with 
composition~$m$, with $m$ given by \eq{eq:op}. We emphasize that $P(m)$ depends 
on all the system parameters, in particular the fugacity $z$, and the system 
size~$L$. Note also that, due to symmetry, $P(m) = P(-m)$, and that this 
symmetry holds irrespective of whether a porous medium is present.

To facilitate a finite-size scaling analysis (both for the bulk system, and 
inside the porous medium) four different system sizes were simulated: 
$L/\sigma=13.57; 17.07; 20.57; 24.07$ (in the figure legends, we report the 
system size rounded down to the nearest integer). For these system sizes, the 
total number of mobile particles ranged between $\sim 1200 - 6000$. The 
simulations were equilibrated for at least $10^5$ GC cycles, and averages were 
obtained following production runs of $10^6-10^7$ GC cycles (longer runs were 
performed for state points close to the critical point). A GC cycle consists of 
a number of attempted MC steps equal to the average total number of particles in 
the system.

For the fluid mixture inside the porous medium, $\rhom>0$, results were 
additionally averaged over at least $M=100$ different configurations of the 
porous medium. The medium configurations were generated by equilibrating a 
system of hard spheres at fixed density $\rhom$ using canonical Monte Carlo 
moves for at least $10^6$ cycles (here a cycle is defined as one attempted move 
per particle; as canonical move we used random displacements of single 
particles). After equilibration, $M$~configurations were collected and stored at 
intervals of $10^5$ cycles. Then, the mobile $AB$ particles of the fluid binary 
mixture were randomly inserted in the hollow cavities of the porous medium, and 
the distribution $P(m)$ of \eq{eq:opd} was obtained in productions runs lasting 
$10^6-10^7$ GC~cycles.

\section{Results}

\subsection{Locating the critical point}

\begin{figure}
\begin{center}
\includegraphics[width=0.75\columnwidth]{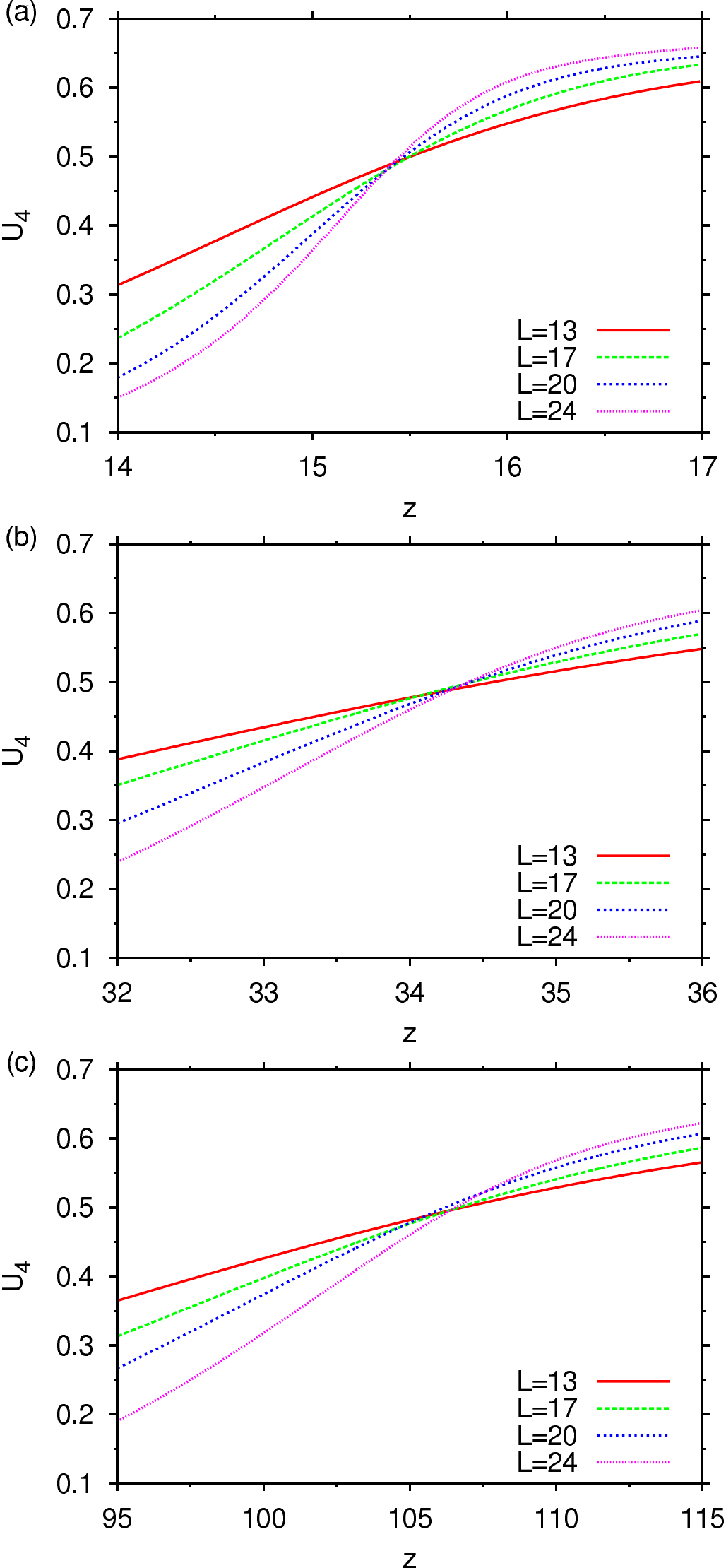}
\caption{\label{fig1} Binder cumulant $U_4$ as a function of the fugacity $z$ 
for different system sizes $L$. The upper panel (a) shows the bulk result. 
Panels (b) and (c) show the result obtained in the presence of the porous 
medium, at medium density $\rhom=0.1$ and 0.2, respectively. The intersection of 
the curves for different $L$ yields the critical fugacity $\zcr$ (\tab{tab2}).}
\end{center}
\end{figure}

\begin{figure}
\begin{center}
\includegraphics[width=0.75\columnwidth]{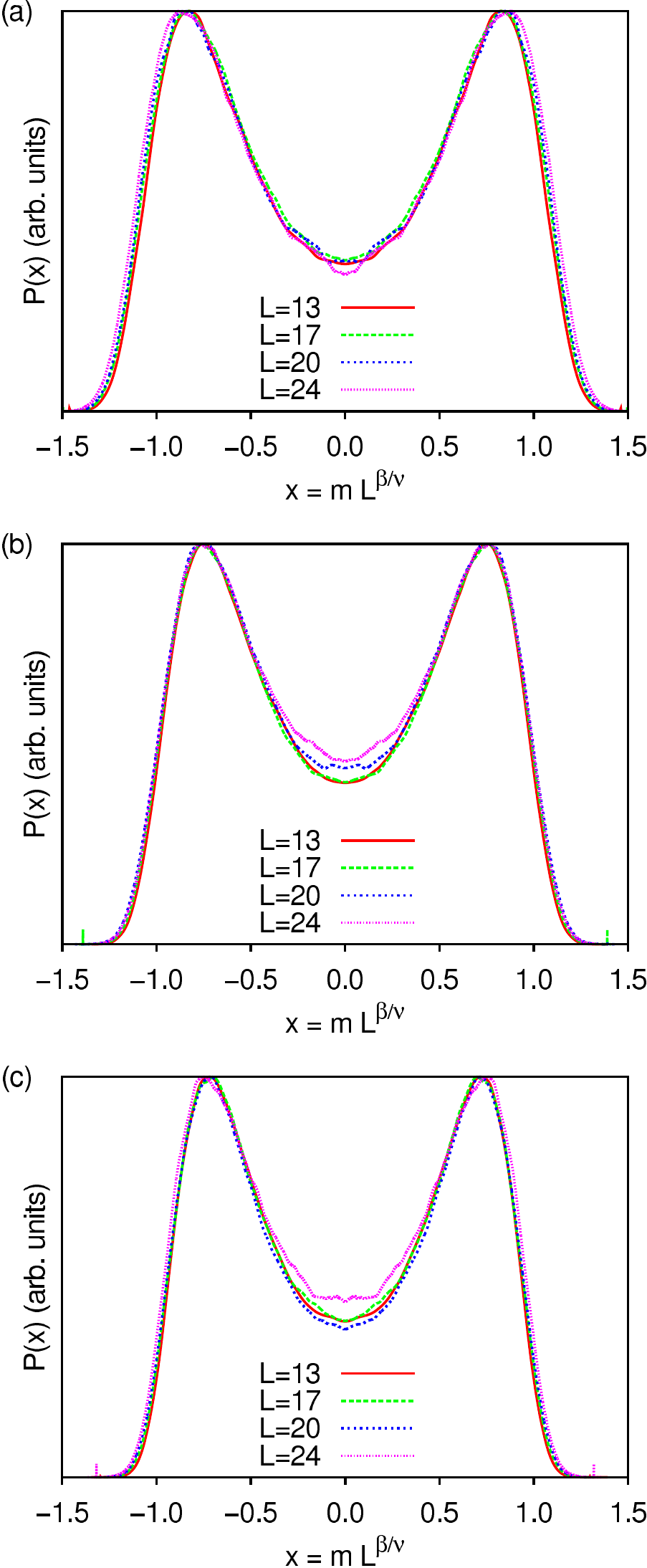}
\caption{\label{fig2} The distribution $P(m)$ obtained at the critical fugacity 
$z=\zcr$ for various system sizes $L$ and scaled conform \eq{eq:sc}. The upper 
panel (a) shows the bulk result ($\rhom=0$) using bulk Ising critical exponents. 
Panels (b) and (c) show the result obtained in the presence of the porous 
medium, at medium density $\rhom=0.1$ and 0.2, respectively, where RIM critical 
exponents were used. Note that, in all panels, the distributions were explicitly 
symmetrized \ahum{by hand} after the simulation had completed.}
\end{center}
\end{figure}

\begin{table}
\centering
\caption{\label{tab1} Critical exponents $\beta$, $\gamma$ and $\nu$ of the 
universality class of the bulk Ising model, and the random Ising model (RIM) 
taken from various references~\cite{pelissetto:2002, hasenbusch.toldin:2007}. 
The spatial dimension $d=3$.}
\ruledtabular
\begin{tabular}{lcccc}
 & $\beta$  &  $\gamma$  & $\nu$ \\\hline
 Ising & 0.326 & 1.237 &  0.630 \\
 RIM   & 0.354 & 1.341 &  0.683 \\
\end{tabular}
\end{table}

Our first aim is to locate the critical point of the transition. To this end, it 
is convenient to consider how the shape of the distribution $P(m)$ changes with 
the fugacity. In the bulk, we recover the behavior typical of a critical 
transition. At high fugacity, $P(m)$ is bimodal with two well-resolved peaks, 
indicating two-phase coexistence. At low fugacity, $P(m)$ is a single peak 
centered around $m=0$. At intermediate fugacities, the system becomes critical, 
where $P(m)$ remains bimodal, but with overlapping peaks. The critical fugacity 
$\zcr$ is obtained via the Binder cumulant
\begin{equation}
\label{eq:u4}
 U_4 = 1 - \frac{ \avg{m^4} }{ 3 \avg{m^2}^2 } ,\quad
 \avg{m^p} = \int_{-\infty}^{\infty} m^p P(m) dm,
\end{equation}
which becomes $L$-independent at the critical point~\cite{binder:1981}. The 
result is shown in \fig{fig1}(a), where $U_4$ is plotted as a function of $z$, 
for various system sizes $L$. The curves strikingly intersect, from which $\zcr$ 
can be accurately extracted (\tab{tab2}). At the critical point, not only the 
cumulant is scale invariant, but in fact the entire distribution 
$P(m)$~\cite{binder:1981, citeulike:4823447}
\begin{equation}
\label{eq:sc}
 z=\zcr: \quad
 P(m) \propto P^\star( a_m L^{\beta/\nu} m),
\end{equation}
with $\beta$ ($\nu$) the critical exponent of the order parameter (correlation 
length), $P^\star(x)$ a scaling function that does not depend on system size, 
and constant $a_m$. The critical exponents, as well as $P^\star(x)$, are 
characteristic of the universality class. We provide numerical estimates of the 
critical exponents for bulk Ising and RIM universality in \tab{tab1}. In 
\fig{fig2}(a), we plot $P(m)$ obtained at criticality, but with the horizontal 
axis scaled conform \eq{eq:sc}, using bulk Ising exponents. We observe that the 
data for different $L$ collapse, consistent with an Ising critical point. 
However, one should regard these observations with some caution, as the critical 
properties of the RIM are very similar. In fact, the exponent ratio $\beta/\nu$ 
is essentially identical between the two classes (and the same holds for 
$\gamma/\nu$, with $\gamma$ the susceptibility critical exponent). Therefore, 
while the data clearly show the presence of a critical point, they do not 
unambiguously identify the universality class (although for the bulk case, there 
is no reason to doubt Ising universality~\cite{gozdz:2003, wilding:2003, 
kofinger.wilding:2006}.

\begin{table}
\centering
\caption{\label{tab2} Critical point properties of the fluid mixture confined to 
a neutral porous medium of density $\rhom$ obtained in this work. Listed are the 
critical fugacity $\zcr$, and the critical density $\rhoc$.}
\ruledtabular
\begin{tabular}{lccc}
$\rhom$ & $\zcr$ & $\rhoc$ \\
\hline
0   & 15.42 & 0.430 \\
0.1 & 34.09 & 0.403 \\
0.2 & 105.6 & 0.379 
\end{tabular}
\end{table}

In the presence of the porous medium, the behavior of $P(m)$ is similar, and a 
critical point can still be identified. The only complication is that results 
must be meaningfully averaged over the $M \ge 100$ medium configurations. In 
contrast to the RFIM~\cite{vink.binder.ea:2008, citeulike:10004339}, we observed 
that the peak positions in $P(m)$ did not fluctuate much between different 
configurations of the porous medium. For this reason, the probability 
distributions were simply averaged to yield the disorder averaged distribution
\begin{equation}
\label{eq:avg}
 [P(m)] \equiv \frac{1}{M} \sum_{i=1}^M P^{(i)}(m),
\end{equation}
where $i$ labels the medium configurations. The cumulant analysis of $[P(m)]$ is 
presented in \fig{fig1}(b) and (c), for $\rhom=0.1$ and 0.2, respectively. We 
again observe that curves for different $L$ intersect, enabling rather accurate 
estimates of $\zcr$ (\tab{tab2}). The scaling of $[P(m)]$ at criticality is 
confirmed in the corresponding panels of \fig{fig2}, where the critical 
exponents of the RIM were used. Again, we emphasize that this analysis 
accurately locates the critical point, but it does not warrant conclusions 
concerning the universality class.

We also estimated the critical density $\rhoc$. To this end, we monitored how 
the density $\rho_L$ varied with the system size $L$, with $\rho_L$ obtained in 
the finite system at the critical fugacity $z=\zcr$. The latter were 
subsequently extrapolated to the thermodynamic limit using $\rhoc - \rho_L 
\propto 1/L$. We thus ignore any singular behavior in $\rhoc$, which is 
justified for our purposes since the shift $\rhoc - \rho_L$ is typically small. 
The resulting estimates of $\rhoc$ are reported in \tab{tab2}. Our estimate of 
the bulk critical density compares well to $\rhoc = 0.4299$ obtained in 
semi-grand canonical simulations~\cite{gozdz:2003}. Note that, while $\zcr$ 
increases with $\rhom$, $\rhoc$ decreases. The increase of $\zcr$ conforms to 
\ahum{Kelvin-like} behavior, i.e.~a suppression of the transition temperature 
upon increasing confinement. The decrease of $\rhoc$ most likely reflects the 
fact that an increasing fraction of space is occupied by the quenched particles. 

\subsection{Correlation length critical exponent}

\begin{figure}
\begin{center}
\includegraphics[width=0.75\columnwidth]{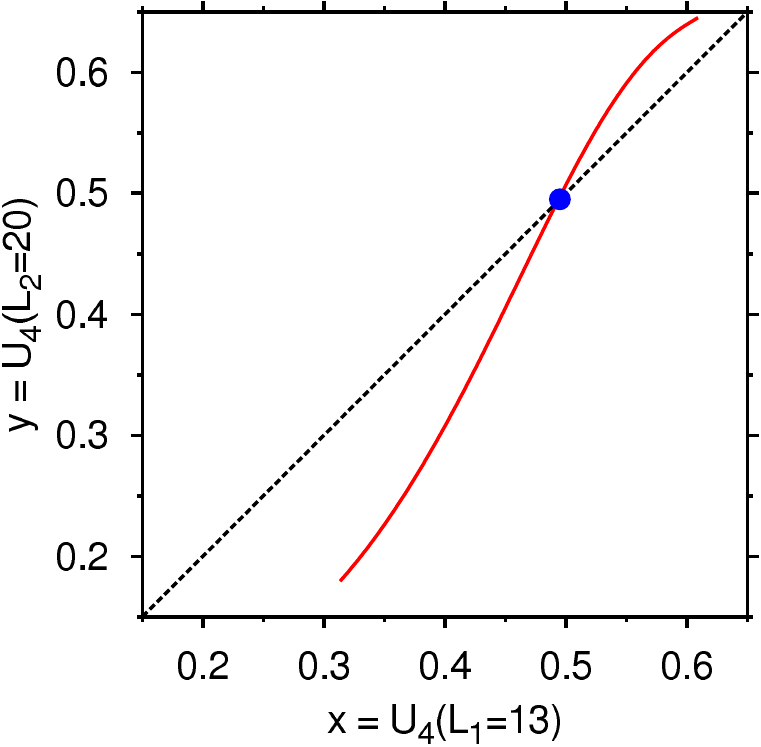}
\caption{\label{demo} Demonstration of the method of \olcite{citeulike:12473115} 
to determine the correlation length critical exponent $\nu$ (data refer to the 
bulk system). The solid curve shows the cumulant of the system with size $L_2$, 
versus the cumulant of the system with size $L_1$. The intersection of this 
curve with the line $y=x$ (dashed) marks the critical point (dot). The slope $s$ 
of the solid curve at the critical point is related to $\nu$ via \eq{eq:demo}.}
\end{center}
\end{figure}

We now attempt to measure the critical exponent $\nu$ of the correlation length, 
using the finite-size scaling approach of \olcite{citeulike:12473115}. To this 
end, we select two of our data sets, corresponding to different system sizes, 
$L_1$ and $L_2$. We then vary the fugacity, and plot the cumulant $y=U_4(L_2)$ 
of the system with size $L_2$ versus $x=U_4(L_1)$ of the system with size $L_1$ 
(the curve is thus parametrized by the fugacity $z$). An example is provided in 
\fig{demo}. The critical point corresponds to the fixed-point condition $ 
U_4(L_1)=U_4(L_2)$, i.e.~where the curve $y(x)$ intersects the line $y=x$ 
(indicated by the dot). The correlation length critical exponent is determined 
by the slope $s=y'(x)$ evaluated at the fixed point
\begin{equation}
\label{eq:demo}
 \nu = \ln b / \ln s, \quad b=L_2/L_1.
\end{equation}
Since $y(x)$ is essentially linear around the fixed-point, the slope $s$ can be 
determined rather accurately. 

\begin{figure}
\begin{center}
\includegraphics[width=0.85\columnwidth]{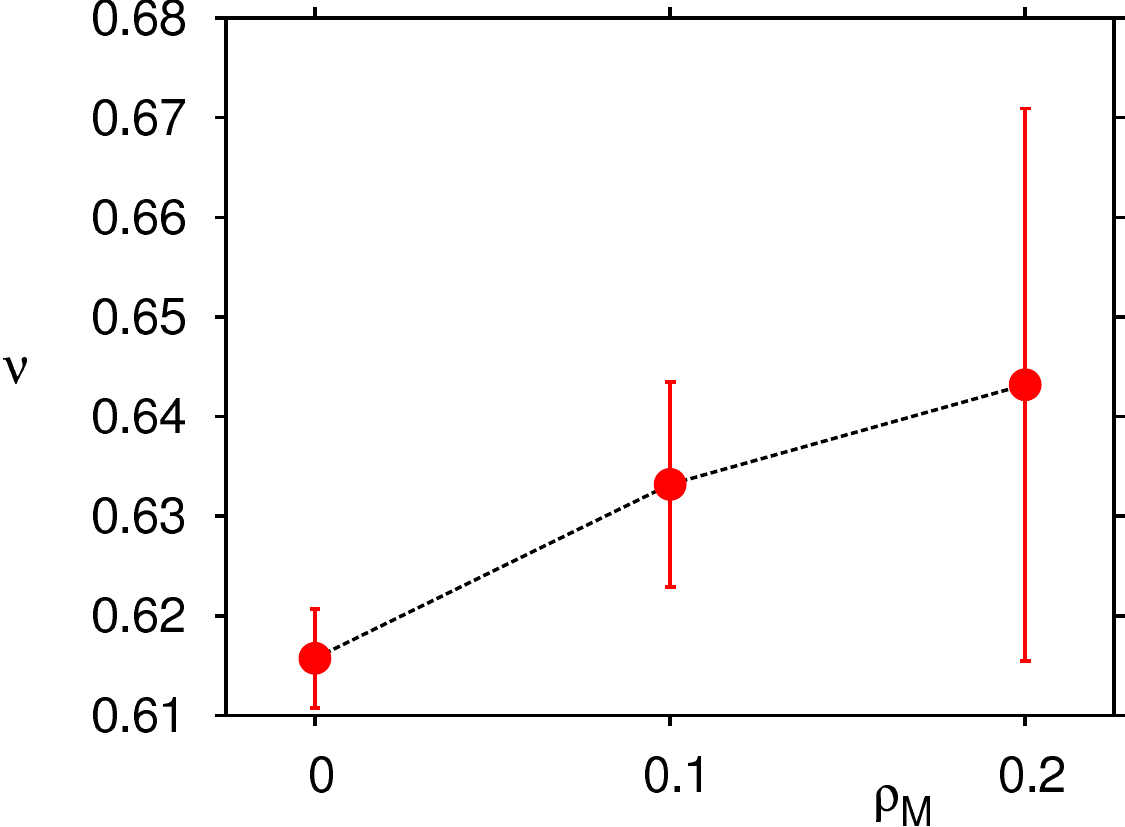}
\caption{\label{fig3} The correlation length critical exponent $\nu$ versus the 
density of the porous medium $\rhom$, as obtained using the method of 
\olcite{citeulike:12473115}. The data reveal that $\nu$ inside the porous medium 
exceeds the bulk value, in qualitative agreement with RIM universality.}
\end{center}
\end{figure}

In \fig{fig3}, we plot the resulting estimates of $\nu$ versus $\rhom$. Since, 
for each value of $\rhom$, we have data for four different system sizes, a total 
of six measurements could be made each time. The dots in \fig{fig3} show the 
average of these measurements, while the error bars reflect the standard error. 
Clearly, the errors are rather large. However, we do observe that $\nu$ inside 
the porous medium ($\rhom>0$) is larger than its bulk ($\rhom=0$) value, a trend 
which is at least qualitatively consistent with RIM universality.

In principle, a similar analysis can also be used to determine the critical 
exponent ratio $\beta/\nu$~\cite{rzysko:2000}. However, as mentioned before, the 
latter is essentially identical for the Ising and RIM universality class, and so 
we did not pursue this.

\subsection{Distribution of pseudo-transition points}

\begin{figure}
\begin{center}
\includegraphics[width=0.75\columnwidth]{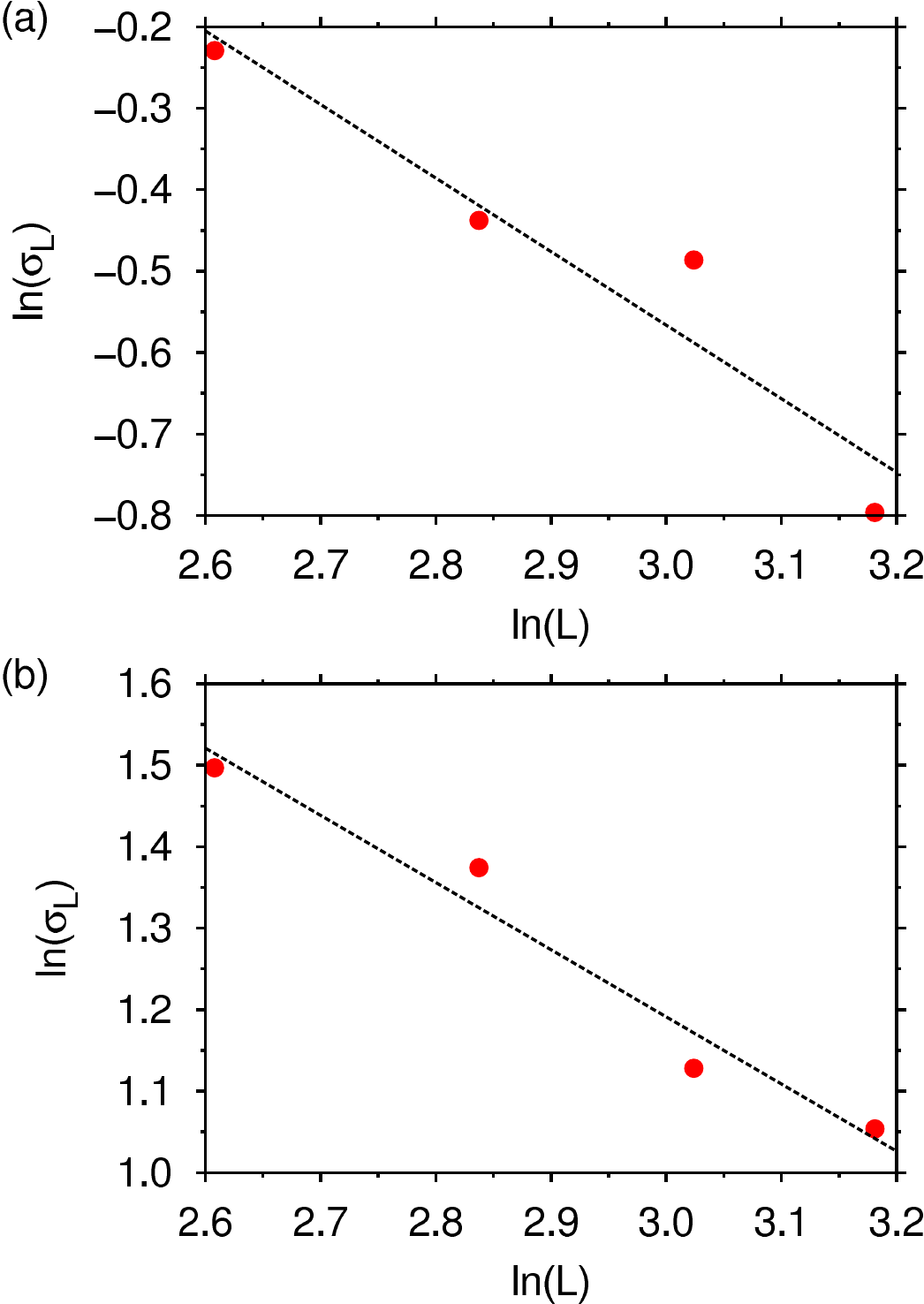}
\caption{\label{fig9} The decay of the fluctuation $\ln \sigma_L$ in the 
pseudo-transition points (defined via the maximum of the susceptibility) as a 
function of $\ln L$, for porous medium densities $\rhom=0.1$ (a) and $\rhom=0.2$ 
(b). The data are approximately linear, indicating a power law decay $\sigma 
\propto 1/L^k$, with $k \sim 0.9$ obtained by fitting (dashed lines).}
\end{center}
\end{figure}

\begin{figure}
\begin{center}
\includegraphics[width=0.75\columnwidth]{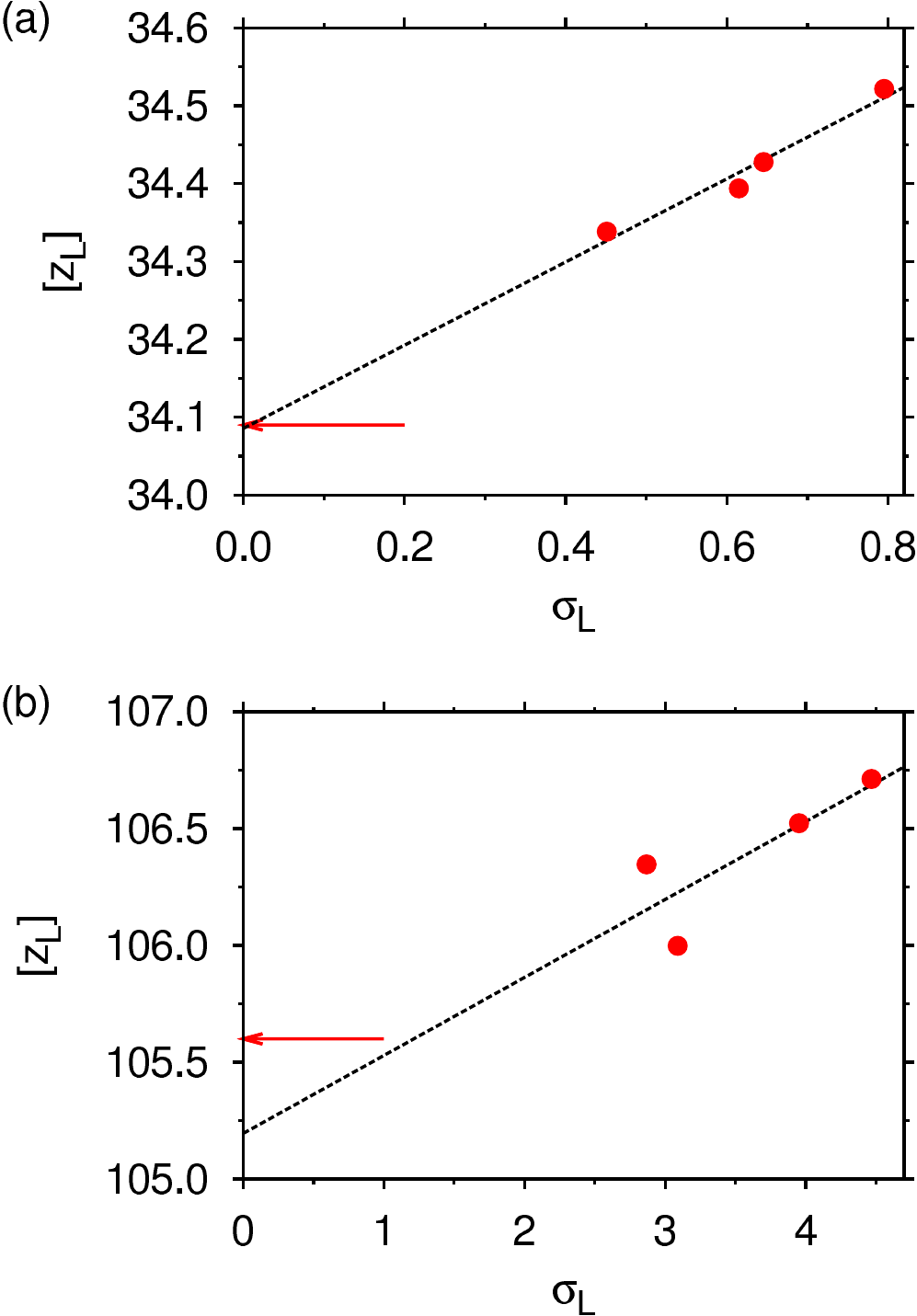}
\caption{\label{inter} Variation of $[z_L]$ with $\sigma_L$, for $\rhom=0.1$ (a) 
and $\rhom=0.2$ (b). The dashed lines are linear fits, whose intercept 
corresponds to $\zcr$. The arrows indicate the estimates of $\zcr$ obtained from 
cumulant intersections.}
\end{center}
\end{figure}

We now consider the distribution of pseudo-transition points; the latter are 
frequently encountered in systems containing quenched disorder, and their 
analysis has attracted much attention~\cite{citeulike:6350198, 
citeulike:6672115, citeulike:6348195, aharony.harris:1996, citeulike:6673387}. 
To be specific, in a finite system of size $L$, the fugacity $z_{L,i}$ where the 
system becomes pseudo-critical, fluctuates between the $i=1,\ldots,M$ 
realizations of the porous medium (the term pseudo-critical is used because a 
finite system never becomes truly critical, of course). The pseudo-critical 
fugacity $z_{L,i}$ may be defined as the fugacity where the susceptibility
\begin{equation}
 \chi_{L,i} = V \left( \avg{m^2} - \avg{|m|}^2 \right),
\end{equation}
reaches its maximum, as measured in the $i$-th realization of the porous medium, 
and with $m$ given by \eq{eq:op}. 

The key question is how the disorder fluctuation
\begin{equation}
 \sigma_L^2 = [z_L^2]-[z_L]^2, \quad
 [z_L^p] = \frac{1}{M} \sum_{i=1}^M z_{L,i}^p,
\end{equation}
decays with the system size $L$. In general, one expects a power law decay: 
$\sigma_L \propto 1/L^k$, with $k>0$. According to the Brout argument $k=d/2$, 
with $d$ the spatial dimension~\cite{citeulike:7522088}. The Brout argument is 
correct, provided the correlation length is finite, such that the system will 
eventually self-average. However, at a critical point, the correlation length is 
infinite, and self-averaging is violated. In that case, the fluctuations decay 
slower, $k=1/\nu$, with $\nu$ the critical exponent of the correlation 
length~\cite{citeulike:6348195, aharony.harris:1996}. Note that, since 
fluctuations may never decay faster than self-averaging, an interesting 
inequality $\nu>2/d$ is implied~\cite{citeulike:7465982, citeulike:10807756}.

In \fig{fig9}, we show how $\sigma_L$ decays with $L$, for both densities of the 
porous medium. Note that a double-logarithmic scale is used. The data are 
compatible with a power law decay. In addition, the exponent of the decay, $k 
\sim 0.9$, is smaller than $d/2=1.5$, showing that self-averaging is violated, 
which is indeed expected for RIM universality. The actual exponent values are, 
however, rather far removed from RIM values (as were our $\nu$ estimates of 
\fig{fig3}). We believe the most likely explanation is the limited number of 
porous medium realizations that we could simulate, and so $\sigma_L$ could not 
be determined very accurately.

The fact that $\sigma_L \propto 1 / L^{1/\nu}$ is also interesting in relation 
to the average pseudo-transition point $[z_L]$, whose shift from its 
thermodynamic limit value $\zcr$ is given by the same form: $\zcr - [z_L] 
\propto 1 / L^{1/\nu}$. Consequently, a graph of $[z_L]$ versus $\sigma_L$ 
should be linear, with the intercept corresponding to $\zcr$. The result is 
shown in \fig{inter}, for both densities of the porous medium. The arrows 
indicate the estimates of $\zcr$ obtained from the cumulant intersections of 
\fig{fig1}. While for $\rhom=0.1$ the agreement between both methods is very 
reasonable, the data for $\rhom=0.2$ reveal significant scatter. Nevertheless, 
the discrepancy remains within 1~\%, and so we conclude that the expected scaling 
is confirmed. \\

\section{Discussion}

In this work, we have considered the critical behavior of a fluid confined to a 
random porous medium consisting of neutral walls. Our aim was to confirm the 
universality class of the corresponding liquid-vapor transition, expected to be 
the one of the random Ising model. While it remains extremely difficult to 
accurately obtain critical exponents for this {\it off-lattice} system, evidence 
of random Ising behavior is revealed by the disorder fluctuations. By monitoring 
the fluctuations in the pseudo-transition temperatures between different 
realizations of the porous medium, clear violations of self-averaging are 
observed. Within the limitations of our data, these disorder fluctuations were 
seen to scale with the system size as would be expected for the random Ising 
model. Also the trend of the critical exponent $\nu$ associated to the divergence
of the correlation length is compatible with the critical behavior of the random
 Ising model. Nevertheless, it is clear that much more computer power would be needed 
to reach the accuracy levels typical of lattice spin 
models~\cite{citeulike:9329724}. We surmise that for such a high-resolution study the disorder
 averages should be calculated over several thousands realizations of the quenched 
disorder, whereas the present study adopted a few hundreds samples only.

For such a possible future study, it is advisable to restrict $\rhom \sim 0.1$ 
or so. This value is large enough to induce random Ising effects, yet small 
enough to avoid the severe equilibration problems that set in at higher medium 
densities. Another quantity that would also be interesting to monitor is the 
coexistence diameter~\cite{vink:2006}. Following the Harris 
criterion~\cite{harris:1974}, the critical exponent of the specific heat 
$\alpha$ is negative for the random Ising model, but positive for the bulk Ising 
model. Such a change in sign might yield a more pronounced numerical signature 
in simulation data.

\acknowledgments

R.V. acknowledges financial support by the German research foundation (Emmy 
Noether grant VI~483). G.P. acknowledges the National Research Foundation (NRF) for financial
 support through Grant No. 80795.

\bibliographystyle{apsrev}
\bibliography{refs_VINK,REFS}

\end{document}